\documentclass[3p,times,twocolumn]{elsarticle}

\usepackage{ecrc}

\volume{00}

\firstpage{1}

\journalname{Nuclear Physics B Proceedings Supplement}

\runauth{A. Yu. Smirnov}

\jid{nuphbp}

\jnltitlelogo{Nuclear Physics B Proceedings Supplement}

\usepackage{amssymb}

\usepackage[figuresright]{rotating}

\newcommand{\be}{\begin{equation}}

\newcommand{\ee}{\end{equation}}

\newcommand{\bea}{\begin{eqnarray}}

\newcommand{\eea}{\end{eqnarray}}

\begin{document}

\begin{frontmatter}




















\author{A. Yu. Smirnov\corref{cor2}\fnref{label1,label2}}

\ead{smirnov@mpi-hd.mpg.de}

\dochead{}


\title{Riddle of the Neutrino Mass}





\address[label1]{Max-Planck-Institute for Nuclear Physics, Saupfercheckweg 1, D-69117 Heidelberg, Germany}

\address[label2]{International Centre for Theoretical Physics, Strada Costiera 11, I-34100 Trieste, Italy}


\address{}

\begin{abstract}

We discuss some known approaches and results as well as few new ideas concerning 
origins and nature of neutrino mass. 
The key issues include (i)  connections of
neutrino and  charged fermions masses, 
relation between masses and mixing,  energy scale of new
physics behind neutrino mass where possibilities spread from the Planck and GUT 
masses down to a sub-eV scale. The data hint  
two different new physics involved in generation of  neutrino mass. Determination 
of the CP phase as well as  mass hierarchy 
can play important role in identification of new physics.
It may happen that sterile neutrinos provide the key
to resolve the riddle.


\end{abstract}


\end{frontmatter}







\section{The riddle}
\label{riddle}


There is something hidden and beyond the standards which
\begin{itemize}

\item 
strongly suppresses, 

\item  
badly confuses,  
and 

\item
violates the law, or maybe, 
doesn't (which is difficult to prove). 

\end{itemize}
And probably the first and the second  are  because of the third.
{\it What is this?}~\footnote{Talk given at the Neutrino Oscillation Workshop, NOW 2014,
Conca Specchiulla (Otranto, Lecce, Italy), September 7 - 14, 2014}

Adapted to physicists this would sound as follows:
What is behind of

1. Smallness of neutrino mass in comparison
to masses of the charged leptons and quarks;  
weak (or no) mass hierarchy of neutrinos. 

2. ``Unusual" lepton mixing  pattern with two large  mixing
angles  (one being close to maximal) and one small which 
differs from the quark mixing;

3. Plausible violation of the lepton number. 

Connected questions (mostly addressed to experiment):   
What is the type of mass spectrum (quasi-degenerate, hierarchical) 
and what is the mass ordering?
What is nature of neutrino masses: 
Dirac versus Majorana,  ``hard'' or ``soft'' ({\it i.e.},  medium dependent)?
Recall that in oscillation experiments
we probe the dispersion relations and not masses immediately.
Effective neutrino masses in oscillation
experiments,  in beta decay, in cosmology and
$\beta \beta$-decay can be different. 

Does the nature of neutrino mass differ
from the nature of the quarks and charged lepton masses?
Indeed, usual neutrino masses can be strongly suppressed,
e.g. by the seesaw, so that ``unusual contributions"  dominate.
Are sterile neutrinos, if exist,  relevant
for the solution of the riddle?

I will discuss  some existing approaches and results  
(see also reviews \cite{reviews}), and present some 
new points. But before that let me challenge the riddle. 
Do we understand and interpret the data \cite{global} correctly?
Are we asking right questions formulating the riddle?
Is whole the story with neutrino mass misleading? 
For instance, 
concerning smallness of neutrino mass:  Is it  normal or special? 

Special: comparing masses within the third fermion generation we have
\be
\frac{m_3}{m_\tau} \approx 3 \cdot 10^{-11}, 
\label{gencomp}
\ee
and similar ratios are for other generations
if neutrino spectrum is hierarchical.

Normal: neutrinos have  no clear generation structure
as well as the  correspondence ``light flavor -  light neutrino mass'',
especially if the mass hierarchy
is inverted or spectrum is quasi-degenerate.
Therefore  comparison (\ref{gencomp}) can be misleading. 
Furthermore, 
\be
\frac{m_3}{m_e} \approx \frac{m_e}{m_t} \approx 3 \cdot 10^{-6}, 
\ee
{\it i.e.}, the same ratio. 
So, neutrino masses can be treated as  a continuation of the mass spectrum 
of charged fermions with certain gap,  
probably due to neutrality of neutrinos. 
This appears even more plausible if 
originally the two heavier neutrinos 
had masses in  the kev and MeV range, but due to some 
(new?) mechanism  were suppressed by 3 - 6 orders of magnitude.

It is not excluded that the correct 
solution of the riddle (or the key to the solution) 
already exists among hundreds of approaches,
models,  mechanisms,  schemes, {\it etc}.
The problem is then to identify the correct solution.
At the same time  something fundamental can be missed.

In what follows I will make an assessments of several  
existing approaches and results in Sec. 2.
Scales and scenarios of new physics will be discussed in Sec. 3.
In Sec. 4 we consider mixing and  CP-violation.
Sterile neutrinos as the key to solve the riddle 
will be mentioned in Sec. 5.  
We conclude with some guesses.

\section{Facts and Arguments}

The most important aspects of the riddle 
include the following. 

1. {\it Leptons and Quarks:} The riddle is formulated as comparison 
of neutrino mass and mixing with masses and mixing of quarks.
There is no solution of the riddle of quark masses.
Can we then solve the neutrino mass riddle?
Do the efforts make sense? Yes, if

(i) neutrino mass generation
and generation of the charged lepton
and quark  masses  are independent. 
Examples of the corresponding mechanisms
include: Higgs triplet \cite{triplet} , Radiative mechanisms \cite{babu,1loop},
Seesaw type  III \cite{ssIII}, {\it etc.}.

(ii)  we  try to explain only  the difference of masses
and mixing of neutrinos and quarks,
and not the whole masses and mixing pattern.

(iii) we still hope (as it was before) that neutrinos
will uncover something simple and insightful
which will allow us  to solve the fermion  mass riddle
in general.\\

{\it 2. Masses and Mixing:} Should the mixing be included in the riddle?
In the quark sector the answer is affirmative:
relation between masses and mixing \cite{gatto}
\be
\sin \theta_c \approx  \sqrt{\frac{m_d}{m_s}} + ...
\ee
exists and Fritzsch (modified) ansatz gives its generalization to
3 generations.  In the lepton sector there is no clear answer  in view of
observed approximate Tri-Bi-Maximal (TBM) mixing \cite{tbm}.
In the residual symmetry approach which explains the mixing \cite{framework},
there is no connection between masses
and mixing (at least in the lowest order). 
Mixing follows from the form invariance of the mass matrices 
independently of mass eigenvalues. 

On the other hand, maximal mixing  can be associated with 
quasi- degenerate mass states. 

{\it 3. Mixing of quarks and leptons:}
Again we have three possibilities. The two mixings  can be

(i) completely related, with the only difference
that originates from the Majorana nature of neutrinos;

(ii)  Partially related,  as it often appears in models with
seesaw type I \cite{seesaw}, quark-lepton unification,  GUT, 
also with the seesaw type II \cite{triplet}.

(iii)  Largely  unrelated, if neutrino masses are generated by Higgs triplet,
radiative mechanisms, seesaw type II and III.

{\it 4. The riddle and the Dark Universe. }
It can be deep connection of the neutrino mass riddle 
with other problems: Dark Energy
and  Dark radiation, baryon asymmetry in the Universe and inflation.  
E.g.,  the same symmetry can be responsible for smallness of the neutrino masses and 
stability of the dark matter particles. 
So,  the solution may come from ``heaven" or from completely  unexpected side.

\section{The riddle and new physics}

Now, especially after first run of LHC we have 
the riddle of new physics:  where it is? 

\subsection{Two types of new physics}

It seems leptons ``know"  about  quark mixing". 
At the same time there is something qualitatively new in the lepton
sector. 
Probably there are two types of new physics behind
neutrino mass and mixing:

1. ``The CKM type new physics'' which  is common for quarks and leptons. 
It is responsible for small quark mixing and hierarchical
structure of the Dirac masses.

2. ``Neutrino new physics" -- an additional structure in the lepton sector, e.g. 
the Majorana mass matrix of the right handed neutrinos which realizes the 
see-saw mechanism. It is responsible for
smallness of neutrino mass and large lepton mixing. 

These two types are different but  
should somehow know about each other.
A counter example: seesaw with degenerate RH neutrinos. 

\subsection{Scales and scenarios of new physics}

The energy scales of proposed new physics behind neutrino mass 
$\Lambda_{NP}$ spread over 28 orders of magnitude:  from the sub-eV up to 
the Planck scale.  Three possibilities 
are motivated somehow:

1. GUT-Planck mass scale  appears as
\be
\Lambda_{NP} = \frac{V_{EW}^2}{m_\nu}.
\ee
It is along with the unification line: 
high scale seesaw,  
$
m_\nu = - m_D^T  M_R^{-1} m_D, 
$
quark-lepton  symmetry (analogy),  GUT. 

Here there are several possibilities:

a) The heaviest RH neutrino has $M_{3} \sim M_{GUT} \sim 10^{16}$ GeV. 
This can be realized in the $3\nu$ context in the presence of mixing.

b) $M_R = (10^{8} - 10^{14})$ GeV, which can be obtained
in the double seesaw mechanism  as $M_{GUT}^2/ M_{Pl}$ \cite{dseesaw}.

Gauge coupling unification, Leptogenesis and probably
BICEP-II  are in favor of this possibility.

c) $M_R = (10^{16} - 10^{18})$ GeV,  
which can be realized if many  ($\sim 10^2$) 
heavy singlets (RH neutrinos) exist as is expected from
string theory \cite{manyrh}. 

The GUT-Planck scale scenario has, however,  the problem. 
The simplest seesaw implies new physical scale
$$
M_R =  m_D^2 /m_\nu \approx 10^{14} {\rm GeV}  \ll M_{Pl}. 
$$
Correction to the Higgs boson mass due to coupling with RH neutrinos 
equals \cite{vissani,stabil} 
$$
\delta m_H^2  = \frac{y^2}{(2\pi)^2} M_R^2 
\log \left(\frac{q}{M_R}\right) \approx 
\frac{M_R^3 m_\nu}{(2\pi V_{EW})^2} \log \left(\frac{q}{M_R}\right), 
$$
where $V_{EW}$ is the electroweak  VEV and $y$ is the Yukawa coupling. 
For usual seesaw with $M_R \sim 10^{14}$ GeV,  
one would get $\delta m_H^2 = (10^{13} {\rm GeV})^2$.  
The straightforward way to solve the problem is to reduce the scale of seesaw mechanism 
so that $M_R \sim 10^{7}$ GeV. This implies small yukawa couplings 
according to equation above: 
\be
y = 2\pi\frac{\delta m_H}{M_R \sqrt{(log(m_H/M_R)}}  y < 2 \cdot 10^{-5} , 
\ee
which in turn requires explanation.
Another problem is that the mass $M_R$ is below
the lower limit from  successful leptogenesis:
$M_R > 10^8$ GeV \cite{leptogen}. 

Possible solution could be some new physics
which leads to cancellation of the $\nu_R$ contribution to $m_H$.
E.g. due to loop with new scalars which have 
the couplings with usual Higgs as sneutrinos in SUSY.  
This is a kind of ``ad hoc
supersymmetry''. Cancellation will be absent at the
two loop level. But this is enough to suppress
the contribution $\delta m^2_H$, so that
$M_R$ can satisfy the leptogenesis bound.
Stronger cancellation at high loop level will require
essentially reconstruction of complete SUSY.
(See also \cite{fabbrichesi}.)

2. The electroweak - LHC  scale: 
\be
\Lambda_{NP} = V_{EW} \div E_{LHC}.
\ee
The lower edge is motivated by already existing scale,  whereas 
the upper one - mostly by logic of ``looking under the lamp".

Here  there is no hierarchy problem (even without SUSY). 
New particles at (0.1 - few) TeV scale are expected which 
can be tested  at LHC.  LFV decays can be at the 
level of sensitivity of  the present experiments. 
The low scale mechanisms include

1. Low scale seesaw, $\nu$MSM \cite{numsm}, low scale LR symmetry model \cite{LRmod},
R-parity violating SUSY with neutralino as RH neutrino \cite{rpv}, 
inverse seesaw with very small lepton violation term.

2. Radiative mechanisms with one, two,  three loops;
high dimensional operators; radiative see-saw. 

3. Small VEV:  Higgs
triplet, new Higgs doublets. 
Some connection to Dark Matter can be realized.  

$\nu$ MSM deserves special attention in view of 
possible (although controversial) observation  of the  astrophysical 
$3.5$ kev X-line and non-observation of new physics at LHC and other experiments. 
In $\nu$MSM  everything is below EW scale,  
and correspondingly, nothing is up to the Planck scale.
This  implies very small neutrino Yukawa couplings.  
The RH sector consists of 
two heavy RH neutrinos of  few 100 MeV - 
GeV mass with extremely  small (below eV) splitting. 
These neutrinos generate masses of active neutrinos via seesaw, 
and  the lepton asymmetry in the Universe 
via oscillations. They  can be produced in  B-decays (BR  $\sim 10^{-10}$ )
\cite{SHIP}.

The third RH neutrino, $\nu_s$,  has mass 
(3 - 10) keV and very small mixing with active neutrinos. It 
composes  the ``cooled'' warm dark matter in the Universe and  
its radiative decays explains the 3.5 kev  photon line.
Higgs inflation can be realized here \cite{bezrukov}.

Several features pose doubts in this minimal scenario: 
in particular,  extremely small splitting of $M_2$ and $M_3$,  
and ``decoupling'' of  $\nu_s$ from generation of the active neutrino masses.  
Indeed, contribution of $\nu_s$ to the masses of active neutrinos
equals 
\be
\delta m_a =  \frac{1}{4} \sin^2 2\theta_s m_s \approx  (3 - 4) \cdot 10^{-7}~{\rm eV}, 
\ee
which is much smaller than 
the smallest  relevant term of the mass matrix:
$\sim \sin \theta_{21} \sqrt{\Delta m_{21}^2} \sim 4 \cdot 10^{-3}$ eV. 
So, essentially this neutrino  decouples from ``seesaw" 
which indicates that $\nu_s$ is not normal RH neutrino. 
It can be that the standard (high scale) seesaw is realized  
with three RH neutrinos and $\nu_s$ is an additional state which 
mixes very weakly with neutrino system.

3. The eV - sub eV scale: 
\be
\Lambda_{NP} \sim  m_\nu,
\ee
that is,  the neutrino mass itself
can be  the fundamental scale of new physics, 
and not just spurious quantity made of some other scales as in 
see-saw. 
This can be related to the dark sector of the Universe, 
dark energy, MAVAN  as possible realization \cite{mavan}, 
existence of new relativistic  (dark radiation).  
It is less explored possibility. 

Very light dark sector  may include
(i) new scalar bosons (majoron,  axions), 
(ii) new fermions (sterile neutrinos, baryonic neutrinos \cite{baryonnu}),
(iii) new gauge bosons (e.g. dark photons) \cite{darkpho}. 
This sector may be related 
to the eV-scale seesaw with RH neutrinos   
for LSND/ MiniBooNE/reactor/Ga anomalies \cite{ev-ss}. 

Tests of such a possibility include
5th force searches experiments; 
searches for modification of dynamics of neutrino oscillations,  
that is,  checks of standard oscillation formulas, {\it etc.}.
 
\section{Mixing and CP-violation}

\subsection{PMNS and CKM}

In a spirit of two types of new physics 
and partial relations of quark and lepton mixing we can 
assume that 
\be
U_{PMNS}  = U_{CKM}^+  U_X , 
\label{pmnsckm}
\ee
where $U_{CKM}$ follows from the charged leptons 
or Dirac matrices of neutrinos in the flavor basis.
This is  the ``CKM type new physics'' 
which generates  
hierarchical structure (similar to $V_{CKM}$)
and determined (as in Wolfenstein parametrization) by powers of
$\lambda \sim  \sin \theta_C$.
$U_X$ comes from new ``neutrino structure''. It
is related to mechanism of neutrino
mass generation which explains smallness of  neutrino mass.
It should be fixed to reproduce correct lepton mixing angles.
Since  $V_{CKM} \approx I$,  $U_X \approx U_{TBM}$.

The prediction from (\ref{pmnsckm}) is  
\be
\theta_{13} \approx \frac{1}{\sqrt{2}} \theta_C. 
\ee
It has been obtained at purely phenomenological level in \cite{tanimoto} 
and in the context of QLC \cite{qlc}.
The prediction is obtained if
\be
U_X = U_{23}(\pi/2)~ U_{12}
\label{ux}
\ee
with $U_{12}$ being arbitrary. 
Maximal (or nearly maximal) 2-3 rotation is needed to explain 
nearly maximal $\nu_\mu - \nu_\tau$ mixing. Special cases
are $U_X = U_{BM}$, which is realized in the QLC \cite{qlc}, 
and $U_X = U_{TBM}$ in the so called TBM-Cabibbo scheme
\cite{tbmcab}.

From  (\ref{pmnsckm}) and (\ref{ux}) we obtain
\be
U_{PMNS}  
\approx U_{12}(\theta_C) U_{23}(\pi/2) U_{12}. 
\label{pmnsckm1}
\ee
To reduce this matrix to the standard form 
one needs to permute
$U_{12}(\theta_C)$  and $U_{23}(\pi/2)$ which leads
to appearance of $U_{13}(\theta_C /\sqrt{2})$.
It gives also small deviation of the 2-3 mixing from maximal one.

It should be stressed however that  the same value of 1-3 mixing
can be obtained in other  ways with completely
different implications.   
Some possibilities are  
\be
\sin^2 \theta_{13} =   
A \frac{\Delta m_{21}^2}{\Delta m_{31}^2}, ~~~~~  
A = {\cal O}(1),   
\ee
which follows from  ``naturalness'' - an absence of fine tuning in  
the mass matrix \cite{zhenia};  
$$
\sin^2 \theta_{13} \approx  \frac{1}{2}\cos^2 2\theta_{23} ~~
{\rm or}~~  \theta_{13} \approx  \sqrt{2} (\pi/4 - \theta_{23})   
$$
from relation between  deviation of the 2-3 mixing from maximal.    
It was predicted in model with   $T^{\prime}$ symmetry
\cite{frampton} but may also follow from the universal 
$\nu_\mu - \nu_\tau$ symmetry violation \cite{myrev}.    
Another interesting relation is  \cite{univrel}
\be
\sin^2 \theta_{13} 
\approx \frac{1}{4} \sin^2 \theta_{12}\sin^2 \theta_{23},  
\ee
which is analogous of  the quark relation 
$V_{ub} = 0.5 V_{us} V_{cb}$. 
This may follow from a kind of Fritzsch ansatz for 
mass matrices (with texture zeros, U(1) symmetry, {\it etc.}).
This implies similar structure of mass matrices of neutrinos and charged fermions 
but with different expansion parameter $\lambda_l$,  furthermore the latter satisfies  
\be
\lambda_l  \approx 1 -  \lambda_q. 
\ee
Expectations  from this scenario are the normal mass hierarchy,
certain relations between masses and mixing;
flavor alignment in the mass matrix.

\subsection{CP-phase prediction}

Let us use the relation (\ref{pmnsckm}) which 
gives correct prediction for 1-3 mixing to get some generic results on 
the CP -phase.  
First, we assume that $U_{CKM}$ is the only source
of CP-violation (similarly to what
happens in the quark sector).
So, there is no CP violation in $U_X$
(similar possibility has been considered previously
in \cite{cp-gen}, \cite{qlc}). 
Then one gets~\cite{basu}
\be
\sin \theta_{13} \sin \delta_{CP} =
(-\cos \theta_{23}) \sin \theta_{13}^q
\sin \delta_q. 
\label{relation}
\ee
Since
$\sin \theta_{13} \sim \lambda$, 
$\sin \theta_{13}^q \sim \lambda^3$
and $\delta_q = 1.2 \pm 0.08$ rad, we obtain from (\ref{relation}) 
\be
\sin \delta_{CP}  \sim \lambda^2 \sim 0.046, 
\ee
or $\delta_{CP} = \delta$ or $\pi + \delta$,  where
\be
\delta \approx 
\frac{\sin \theta_{13}^q}{\sin \theta_{13}}
\cos \theta_{23} \sin \delta_q.
\ee
Thus,  if leptons have the same origin of CP-violation
as quarks, the leptonic
CP violation phase is small (unobservable) 
or very close to $\pi$
which can  be  observed in atmospheric neutrinos. 

There are two implications of this result:

1). If future measurements show that the phase $\delta_{CP}$
deviates substantially from 0 or  $\pi$,
new sources  of CPV beyond CKM should exist 
(e.g. from the RH neutrino sector), or another framework is realized.

2). New sources may have specific symmetries or structures which
lead to particular values of $\delta_{CP}$,
e.g.  $-\pi/2$. Then  
CP from the LH rotation which diagonalizes the
Dirac  mass matrix gives just small corrections.

In general, the phase $\delta_{CP}$ can be large.
Neglecting terms of the order $\sim  \lambda^3$
we obtain \cite{basu}
$$
\sin \delta_{CP} = \frac{
\sin (\alpha_\mu + \delta_x)V_{ud} |X_{e3}| -
\sin \alpha_e |V_{cd}||X_{\mu 3}|}{s_{13}}, 
$$
where $\alpha_\mu$, $\delta_x$ and $\alpha_e$
are parameters of the RH neutrinos.
Some special values of $\delta_{CP}$ can be
obtained under certain assumptions.
If, e.g., $X_{e3} = 0$,  we have
$\sin \delta_{CP} \approx - \sin \alpha_e$.
Furthermore, if  $\alpha_e = \pi/2$, then
$\delta_{CP} \approx 3\pi/2$.
One can easily find structure of the RH neutrino mass matrix 
which leads to these equalities.

In the seesaw type-I  $U_X$
is the matrix which diagonalizes~\cite{basu}
\be
M_X  = - m_D^{diag} U_R^+ (M_R)^{-1}
U_R^* m_D^{diag} . 
\label{mxmat}
\ee
Here  $m_D = U_L m_D^{diag} U_R^+$, and
we assume that $m_D^\nu \sim  m_D^q$.

In contrast to quarks for the Majorana neutrinos
the RH rotation that diagonalizes  $m_D$ becomes
relevant and contributes to $\delta_{CP}$. 
Important special case is model with L-R symmetry.
In the L-R  symmetric basis $U_X = U_R U_S$ 
with  $U_S$ being the matrix which comes from seesaw. 
Due to  the  L-R symmetry:
$
U_R = U_L \sim  V_{CKM}^*  
$, 
and we  assume that there is no CP violation in $M_R$. 
CP violation in $U_R$ is small. However, 
it turns out that seesaw itself can enhance this 
small CPV effect, so that resulting  phase
in the PMNS matrix is large~\cite{basu}. The seesaw enhancement
of the CP violation is related to strong hierarchy of the
mass eigenvalues of $m_D$.


\section{Steriles and neutrino portal}

Effect of different sterile neutrinos on the $3\nu$
structure can be parametrized  as  
\be
m_\nu =  m_a  + \delta m_a, 
\ee
where the first term is the original active neutrino mass
matrix, e.g., from see-saw, 
whereas the second one is the induced mass matrix
due to mixing with steriles. Notice that
$m_a  =  0.025$ eV in the
case of hierarchical mass spectrum.

Three cases are phenomenologically motivated 
which correspond to different mass scales: 

1). The kev mass scale sterile ($\nu$MSM): $\delta m_a \ll m_a$.
Sterile neutrino decouples from generation
of the light neutrino masses, as we said above.  

2). The eV mass sterile with mixing required by 
LSND/MiniBooNE results in  $\delta m_a \sim m_a$. 
This is not a small perturbation,
$\delta m_a$ can change structure (symmetries)
of the original mass matrix completely. In general,
it can be origin of difference of $U_{PMNS}$ and $U_{CKM}$.

3). meV steriles: $\delta m \ll m_a$ \cite{pedro}:
again it can be considered as very small 
perturbation of the $3\nu$ system.

Is $\nu_R$ the key to the solution of the riddle?
Various issues we have discussed here 
(neutrino new physics, scales, symmetries) are related in one way or another 
to  sterile neutrinos. 

In general, we can wright effective neutrino interactions as
\be
\frac{1}{\Lambda^{n(F) -3/2}} L H F,
\ee
where $H$ is the Higgs doublet,  $F$ is the fermionic operator, which is singlet of the SM 
symmetry group, and $n(F)$ is the dimension of $F$. 
Through this ``portal''  neutrinos get  mass and new physics may show up.


\section{Instead of conclusion}

Warning: Formulation of the riddle here may be misleading and we may misinterpret the data. 

Some guesses: Two different types of new physics are involved
in explanation of data:  the CKM type  common
to quarks and leptons and  physics responsible
for smallness of neutrino mass and
large lepton mixing. 
It makes sense to identify the second
one,  which explains the difference between
quarks and leptons masses and mixing. 
Still generation of quark and neutrino
masses can be essentially independent.

New physics at 

- high (GUT) scale: still  appealing;

- EW scale: wait and see LHC14 results;

- sub eV - eV scale: interesting and  
worth to explore further.

New neutrino physics may have certain symmetries
which leads to specific  values of mixing
angles and CP phase.  The CP-phase from the CKM
part is strongly suppressed.

Sterile neutrinos (in one way or another) may turn out to be the key 
to the solution of the  riddle.















\nocite{*}

\bibliographystyle{elsarticle-num}

\bibliography{martin}




\end{document}